# Augmented Tree-based Routing Protocol for Scalable Ad Hoc Networks

Marcello Caleffi, Giancarlo Ferraiuolo, and Luigi Paura

*Abstract*—In *ad hoc* networks scalability is a critical requirement if these technologies have to reach their full potential. Most of the proposed routing protocols do not operate efficiently with networks of more than a few hundred nodes. In this paper, we propose an augmented tree-based address space structure and a hierarchical multi-path routing protocol, referred to as Augmented Tree-based Routing (ATR), which utilizes such a structure in order to solve the scalability problem and to gain good resilience against node failure/mobility and link congestion/instability. Simulation results and performance comparisons with existing protocols substantiate the effectiveness of the ATR.

## I. INTRODUCTION

In the last ten years, Mobile *Ad hoc* NETwork (MANET) technologies are tremendously growing. Most of the research has mainly regarded relatively small networks and has been focused on performances and power consumption related issues. More recently, due to the importance of ad hoc paradigm in applications involving a large population of mobile stations interconnected by a multi-hop wireless network [1], great attention has been devoted to self-organizing routing protocols with satisfactory scalability requirements since most of the available routing protocols operate satisfactorily only up to few hundred nodes [2, 3].

Such protocols don't scale efficiently mainly because they have been derived by modifying traditional routing algorithms, conceived for fixed networks, to cope with the dynamic topology of MANET [4, 5]. More specifically, traditional routing procedures are based on the assumption that node identity equals node routing address (i.e. they exploit static addressing schemes), which of course is not yet valid in MANET scenarios.

Recently, routing protocols have exploited the idea of decoupling identification from location, by resorting to Distributed Hash Tables (DHTs), which are used to distribute node's location information throughout the topology. The information stored in the DHTs is a dynamic network address, which reflects the node topological position inside the network. When the dynamic address of a node has been retrieved from the DHTs by the lookup procedure, the routing procedures [6-8, 11, 12] resort to the topological information present in the address, resembling the routing procedure utilized by wired networks.

According to such dynamic addressing approach, the schemes [6-8, 12] are hierarchically organized exploiting a tree structure for the address space management and routing. Although this structure offers a simple and manageable procedure, it lacks for robustness against mobility and exhibits unsatisfactory route selection flexibility. Routing protocols based on such addressing schemes can determine low performance and poor resilience to node failure/mobility [11]. In order to face with the problem of the incomplete information embedded in the tree-based addressing scheme, we propose to augment the tree structure by storing additional information in the node routing tables. This information is simply acquired by each single node by using the underlying neighbour discovering procedure. The advantage of the proposed routing protocol, referred to as Augmented Tree-based Routing (ATR) protocol, lies in the richer topology knowledge that allows one to resort to multi-path routing.

The outline of the paper is the following: Section II introduces the preliminaries of the proposed protocol, whereas in Section III we describe it in detail. Then, in Section IV, an extensive simulation performance comparison between ATR and other routing protocols is presented. Finally, in Section V, conclusions are drawn.

## II. PRELIMINARIES TO ATR PROTOCOL

The address space structure exploited by the proposed protocol can be represented as a binary tree of $l+1$ levels, where $l$ is the number of bits used for an address (Fig. 1).

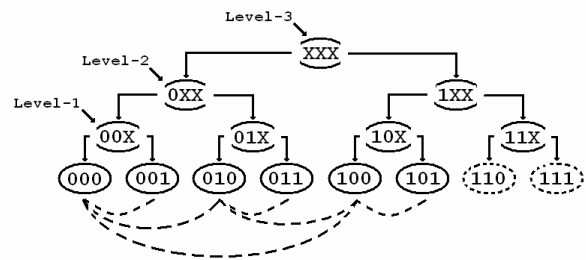

Figure 1 – Address space structure

The leafs represent the network addresses, whereas the physical connections are represented by dotted lines and they do not necessarily correspond to the branches of the address

All the authors are with the DIET, University of Naples "Federico II", via Claudio 21, Naples (Italy), tel:+39-(0)-81-7683810, fax:+39-(0)81-7683149, emails: {name.surname}@unina.it.
This work is partially supported both by National project Wireless 8O2.16 Multi-antenna mEsh Networks (WOMEN) under grant number 2005093248 and by the Italian Ministry of University (MIUR) project S.Co.P.E..

space tree. In such structure, a level-*k* subtree is a set of nodes sharing an address prefix of (*l-k*) bits: a level-0 subtree is a leaf node, where [0xx] is a level-2 subtree containing the four addresses [000], [001], [010] and [011]. A level-*k* sibling of a given address is defined as the subtree that shares the same immediate parent of the level-*k* subtree of the considered address. Such an address allocation structure, exploited in the Dynamic Address RouTing (DART) protocol [12], assures that nodes, whose routing addresses share the same prefix, form a connected sub-graph in the network topology. In particular, the longer the shared address prefix between two nodes, the shorter the expected routing distance in the network topology. In few words, a *tree-base logical structure* on the address space based on connectivity between nodes is introduced and exploited by routing since it appears to be suitable for its hierarchical characteristic. However, this tree-based structure has a low fault-tolerance, since it exists only one path between a node and a sub-set of destinations, i.e. a sibling. The failure of a next hop breaks the connectivity of the network, leaving the destination set disconnected from the node. Another major weakness is that this structure suffers from traffic congestion. This is due to the availability of a unique next hop as a gateway for a whole sibling, i.e. a set of destination nodes, which can be constituted by many nodes.

To overcome such drawbacks, in this paper we propose to *augment* the tree structure by adding redundant paths for the packet forwarding. More specifically, unlike DART protocol in which each node maintains only one possible *next hop* toward the final destination (defining a single path along the tree structure of the address space), in ATR each node maintains and explores all the possible paths through its neighbours to reach the final destination. This is equivalent to use an *augmented tree* structure to perform forwarding, which slightly increases cost, as shown in Section IV. Moreover, the richer network-topology knowledge is exploited to implement *temporal* multi-path strategies, which guarantee better performance and a higher *reliability* [16].

```
0 1 0 0 0 0 1 0     0 1 0 0 0 0 1 0     0 1 0 0 0 0 1 0
1 0 0 0 1 1 1 1     1 0 0 0 1 1 0 1     1 0 0 0 1 1 1 1
0 0 0 1 1 0 1 0     0 0 0 1 1 0 1 0     0 0 0 1 1 0 1 0
0 0 1 0 1 0 1 0     0 0 1 0 1 0 1 0     0 0 1 0 1 0 1 0
0 1 1 1 0 1 1 1     0 1 0 0 0 0 1 0     0 1 1 1 0 1 1 1
0 1 0 0 1 0 1 1     0 1 0 0 1 0 1 0     0 1 0 0 1 0 1 1
1 1 1 1 1 1 0 1     1 1 1 0 0 0 0 0     1 1 1 1 1 1 0 1
0 1 0 0 1 1 1 0     0 1 0 0 1 1 1 0     0 1 0 0 1 1 1 0
```
Physical adjacency matrix     DART logical adjacency matrix     ATR logical adjacency matrix

**Figure 2 – Adjacency matrix for 8 nodes network**

To understand the potentiality of the proposed method, in Fig. 2 we have represented the adjacency matrixes associated with the physical and overlay graphs referring to a network with eight nodes. These matrixes differ for their numbers of '1' (communication links). The first one on the left represents the physical graph, i.e. the graph in which the edge $e_{ij}$ is present if a physical communication link is available between the nodes *i* and *j*. The other two matrixes represent the overlay graphs built upon the physical network by DART and ATR protocols, respectively. The lack of ten edges ('1') in the DART matrix, with respect to the physical and ATR ones, evidences the inability of shortest-path routing protocols, as DART, to build a complete topological view of the network.

As regards the routing issues, Fig. 3 shows the overlay graphs associated with the different path discovery results (DART and ATR) for a full mesh network with four nodes. The graphs show the paths from each node towards two destinations, say node '2' and '4'. The graphs evidence the presence of multiple paths towards the same destination in ATR. Moreover, they show that hierarchical single-path routing protocol does not always provide the shortest path, neither when the network is very simple.

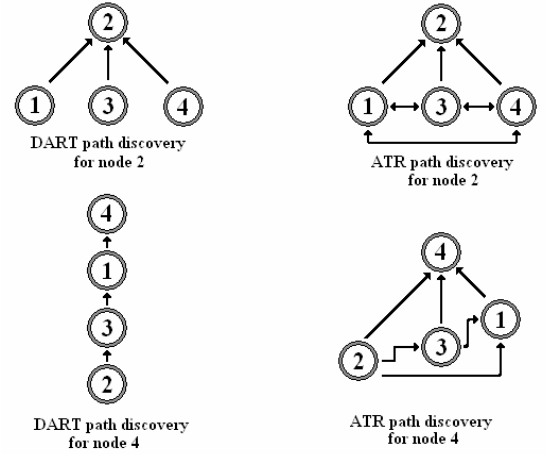

**Figure 3 – Graphs referring to path discovery process**

### III. AUGMENTED TREE-BASED ROUTING PROTOCOL

We distinguish four primary processes in ATR protocol. The *Path Discovery Process* updates the routing table of each node with the routing update sent by neighbour nodes in the *hello* packets. The *Packet Forwarding Process* singles out the right path to route the packets towards the destination. The *Address Allocation Process* selects a network address that reflects the node topological position inside the network. Finally, the *Address Lookup Process* provides the mapping between the unique node identifier used by higher levels and the transient network address used by the Path Discovery and Packet Forwarding processes.

#### A. Path Discovery Process

The proposed path discovery process is a multi-path version of the proactive shortest-path distance-vector one adopted by DART. It takes advantage of the augmented tree structure previously introduced to reduce the effects of node mobility and wireless propagation instability, without increasing the routing overhead.

The Path Discovery Process uses the locally broadcasted *hello* packets exchanged by neighbour nodes to build up and update the routing tables. ATR differs from DART in the number of entries to reach each sibling stored in the routing tables. DART stores a unique entry for each sibling, the shortest one that the node could find. Differently, ATR stores all the available paths that the node could find towards the same sibling. In this way, if the shortest path is unavailable due to mobility, congestion or wireless propagation instability, the node could immediately route the data packets along one of the multiple available paths.

**Table 1 – Routing update of node 1 (address [000])**

| level | sibling | NID | cost | routeLog |
|---|---|---|---|---|
| 0 | 001 | 3 | 1 | 001 |
| 1 | 01X | 2 | 1 | 010 |
| 2 | 1XX | 1 | 1 | 100 |

Table 1 shows an example of the routing update broadcasted by the node '3' with network address [000] in the network of Fig. 3. This routing update advices neighbour nodes only about which destination siblings the sending node could forward packets to, but it does not give information concerning the specific path the packets will be forwarded along. The routing updates need to store only binary information: "*There is no route*" or "*There is at least one route*", so that the routing overhead is the same for both in DART and ATR, as shown by the results concerning the routing overhead in Section IV.B. In Table 1, the Network ID (*NID*) is the identifier of the sibling, i.e. the lowest *id* present in the sibling. It is used by the Address Allocation Process (see Section III.C) to detect the address-duplication event. The cost is self-explanatory and in this work we adopt a simple hop count metric. The *routeLog* is used by the loop avoidance mechanism [12] to discard a route updating, already received. If a node does not receive any *hello* packets from a neighbour in a certain number of update periods, the expired routing update of that neighbour is discarded from the routing table.

**Table 2 – DART routing table of node 4 (address [001])**

| level | sibling | nextHop | ID | cost |
|---|---|---|---|---|
| 0 | 000 | 000 | 1 | 1 |
| 1 | 01X | 010 | 3 | 1 |
| 2 | 1XX | 100 | 2 | 1 |

Let's describe how a node updates its routing table. Suppose that the node '4' with network address [001] (Fig. 3) receives the routing update from node '1', as shown in Table 1. First, node '4' adds an entry for the sibling the address '000' belongs to, i.e. the level-0 sibling [000], with a one-hop cost and node '1' as next hop. Then it looks if the neighbour could act as forwarder for the higher level-k siblings, inspecting if the corresponding *cost* of the routing update has a finite value. In this example the node '4' adds node '000' as forwarder towards the level-1 sibling [01X] and the level-2 sibling [1XX], both with cost 2.

**Table 3 – ATR routing table of node 4 (address [001])**

| level | sibling | nextHop | ID | cost |
|---|---|---|---|---|
| 0 | 000 | 000 | 1 | 1 |
| 1 | 01X | 000 | 1 | 2 |
|   |     | 010 | 3 | 1 |
| 2 | 1XX | 000 | 1 | 2 |
|   |     | 010 | 2 | 2 |
|   |     | 100 | 2 | 1 |

In Table 2 we report the routing table of node '4', for the network of Fig. 3, built by DART protocol, while Table 3 shows the same routing table built by ATR protocol. Also when the network size is very small, only four nodes, the ATR Path Discovery Process can take advantages of multiple neighbours in order to forward packets, thanks to its multi-path approach and its augmented tree-based address-space structure.

*B. Packet Forwarding Process*

The ATR multi-path routing exhibits temporal diversity, i.e. the Path Discovery Process performs a pre-emptive route discovery before the occurrence of route errors. Moreover, ATR could be easily extended to split a data transfer on multiple paths in the spatial domain, to reduce congestion effects and end-to-end delay.

Let us describe the proposed Packet Forwarding Process. According to Table 3, if node '4' with network address [001] must forward a data packet to a node with network address [010], it first looks the entries related to the sibling the destination network address belongs to, i.e. the level-1 sibling [01X]. In this case there are two entries in the routing table, so node '4' will pick up the one exhibiting the least hop count metric, i.e. the node [010]. Otherwise, if there are no entries for the level-1 sibling, node '4' will expand its search to higher sibling, i.e. level-2 sibling [1XX].

Moreover, we take advantage of multi-path defining a cross layer solution to handle with link failures. If a node detects a link failure after the forwarding of a data packet, namely if it does not receive the acknowledgement, the previously used next hop is invalidated. Then the data packet will be re-forwarded using a different path already discovered by the Path Discovery Process. Evidently this leads to higher delays in packet delivery, however it is often more convenient to wait a little more instead of wasting the resources used up to here in packet forwarding [2]. The use of this link-breakage detection technique is another difference of the proposed approach with respect to DART.

*C. Address Allocation Protocol*

ATR protocol makes use of the same address allocation presented in [12], which is a distributed *stateful* approach [14, 15] based on multiple disjoint allocation tables. In few words, when a node joins a network and selects an address, it keeps also the control over a subset of the address space, i.e. a sibling. Nodes exchange information about the utilized addresses and perform both the network-merging event detection and the partition one by locally broadcasting the *hello* packets. Here, we point out only our following change to address allocation process utilized by DART protocol [12], which allows to solve the following issue present in the original address selection procedure. When a node joins a network, it must choose a neighbour to get a valid network address. DART protocol suggests to choose the neighbour with the largest unoccupied address space, i.e. the highest free level-k sibling, to balance the routing table size among nodes. This procedure, as shown by simulation results (Section IV.C), never converges also for small networks since the routing table of the selected neighbour could be not update, and, therefore, the joining node could pick up an invalid address.

Our proposal solves this issue by using as metric for the neighbour selection both the free address-space criteria and the node identifier. Moreover, if an invalid address is acquired from the first selected neighbour, the ATR address selection procedure scrolls the set of neighbours until a valid address will be obtained.

*D. Address Lookup Process*

The Address Lookup Process is built upon a DHT [22]. Unlike DART, our proposal makes use of caching techniques to reduce the delay and the overhead due to the procedure of looking up a network address of a node starting from its identifier. We want to underline that the proposed caching technique could be also used to provide fault tolerance to the whole process. Moreover, we investigate the issue of finding a good hash function, i.e. a hash one that balances the lookup traffic among nodes, with respect to the adopted address allocation procedure.

Every node is part of the DHT system, storing a subset of pairs *<identifier, network address>*. Which pair a node must store depends on the hash function. Let's make an example: suppose the node with identifier $id_1$ joins the network and picks up the network address $add_1$. Then, it will send a *Network Address Update* (NAUP) packet to the node whose network address is equal to $add_3$=hash($id_1$), i.e. the network address [111], where *hash* is a globally known function. Every node belonging to the route path of the NAUP packet will cache the pair $<id_1, add_1>$. If there is no node with network address [111] in the network, the NAUP packet will be routed to the node having the network address with the bigger common prefix with [111], i.e. the address [110] or [100]. When a node, i.e. the node $id_2$ with address $add_2$, must send a packet to node $id_3$, it will send a *Network Address Request* (NARQ) packet to the node with network address $add_3$=hash($id_1$), i.e. [111]. If $add_3$ is not used, the route process is the same as for NAUP packets. Every node belonging to the route path of the NARQ packet will cache the pair $<id_2, add_2>$ and, if it caches the request pair, it will immediately reply to the request. Differently, it will be the node with network address $add_3$ and identifier $id_3$ that will reply to the request with a *Network Address Reply* (NARP). Every node belonging to the route path of the NARP packet will cache the pairs $<id_2,add_2>$, $<id_1,add_1>$ and $<id_3,add_3>$. Moreover, the *hello* packets are also used to broad the pairs among neighbours for caching.

Owing to the adopted address allocation procedure, the network addresses are not assigned with the same occurrence. For example, the address with all zeros is more frequent than the address with all ones. Therefore, we propose a simple yet effective hash function that tries to uniformly distribute the pairs among the nodes whenever the network addresses are not all employed. More specifically, our proposed hash function operates as follows: the returned network address is the binary representation of the id, reversed in the order of the symbols and, finally, with the most significant bit flipped.

## IV. SIMULATION RESULTS

In this section, we present a numerical performance analysis of the proposed routing protocol, resorting to *ns-2* (version 2.30) network simulator [21]. We adopt the standard values for both the physical and the link layer to simulate an IEEE 802.11a Lucent network interface with *Two-Ray Ground* as channel model. The duration of simulation experiment is set to 750 seconds, while the sizes of the scenario areas are chosen to keep the node density equal to 64 nodes/Km$_2$. This value corresponds to a mean node connectivity degree of 12, which is a reasonable value to avoid the presence of isolated nodes [17]. To generate mobile topologies, we have adopted the *Random Way-Point* as mobility model [18]. Since proactive routing protocols are not suitable for networks with very high levels of mobility, the mobility parameters have been set to simulate moderate mobility; specifically, the speed values are uniformly taken in the [0.5m/s; 5m/s] range, and the pause times uniformly taken in [0s; 100s]. Our comparison does not include the network address lookup layer, which is replaced with a global lookup table available to all nodes. However in order to develop the proposed address lookup process, and in particular to adopt effective caching techniques, we have extensively investigate the overhead due to the network address lookup and update functions, and here we report a subset of the simulation results.

*A. Path Discover Process*

As already explained, the proposed multi-path approach has no effect on the routing overhead: both the size and the rate of the *hello* packets are the same for DART and ATR protocols. Instead, the node memory requirements are not the same: ATR protocol requires that nodes store all the available paths towards each sibling.

In this subsection, we evaluate the memory requirements of ATR and compare them with the ones of DART, in terms of routing table size. We have run a set of trials to measure the average number of routing tables entries of all participating nodes. As shown in Fig. 4, ATR exhibits stronger memory requirements. However, as the number of nodes grows, the number of routing table entries saturates: this confirms that the proposed augmented tree-based address space structure scales satisfactorily.

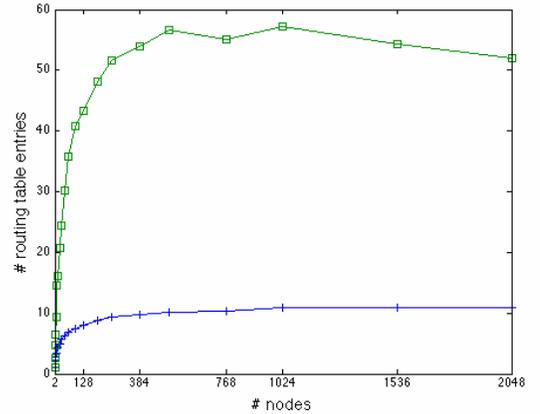

**Figure 4 - Memory requirements comparison**

Clearly, the routing table size depends on the node density. Different node densities have been experimented, confirming the same general trend.

*B. Packet Forwarding Process*

Neither DART nor ATR were designed to optimize the throughput. In fact, their main requirement is to achieve scalability. Moreover, they are lacking in optimization work behind more widespread protocols. Nevertheless, we are interested in comparing ATR not only with DART, but also with other two popular routing protocols, AODV [9] and DSR

[10]. Let us underline that ATR does not adopt spatial diversity multi-path routing, so the comparison with shortest-path protocols makes sense.

In comparing these protocols, we choose to evaluate them using the following metrics: *packet delivery ratio*, *path stretch* measured in number of hops, and *routing overhead*. The data traffic is modeled as CBR flows over UDP protocol. We do not adopt the TCP as transport protocol to avoid the effects of elasticity of TCP flow control on routing performances [19]. The data pattern is the most common one in simulation for ad hoc networks: the Random Traffic Model. The global load offered is kept constant at 250Kb/s, in order to avoid running out of capacity due to multi-hop approach. The node load offered scales as $O(1/n)$ to simulate sustainable data traffic, taking in account the routing overhead [20]. Each flow has a start- and end- time uniformly picked in [450s, 720s], in order to achieve the address allocation convergence before data forwarding can be performed (Section IV.C), and to guarantee a 30 seconds cold-down period to complete data packets delivery.

The packet delivery ratio describes the loss rate that will be seen by upper layers protocols. Fig. 5 (and the following in this subsection) shows this metric normalized to the packet delivery ratio of the ATR protocol. The experimental results show that the ATR scales always better than DART. Regarding to reactive protocols, when the number of nodes is relatively small, they perform well. However, the performances of ATR remain comparable with respect to AODV and DSR ones also in such situation. Differently, when the number of nodes grows, the reactive protocols lose their initial performance advantage and ATR outperforms. This trend is earlier manifested by DSR that exhibits the worst performance.

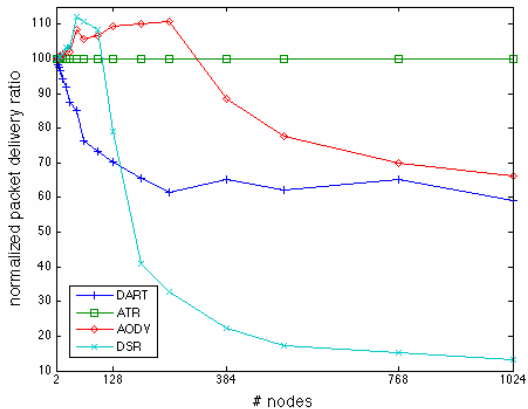

Figure 5 – Normalized packet delivery ratio

Fig. 6 shows the mean path stretch (that is the number of hops a packet spends to reach its destination) normalized to the one of ATR protocol. In absence of congestion, the path stretch measures the ability of a routing protocol to efficiently use network resources by selecting the shortest path to reach the destination. Clearly, the ATR multi-path property leads to longer paths with respect to shortest-path protocols, as illustrated in Fig. 6. However, it is important to note that this metric is evaluated on the base of the correctly delivered packets. When the number of node grows, ATR delivers more packets with respect to the other protocols; then it is reasonable to assume that the packets delivered by ATR, and missed by the other ones, experience longer paths. This is also confirmed by comparing DART and ATR performances for small networks. When both the protocols reach the same packet delivery ratio values, ATR is able to find paths shorter of about 65% with respect to DART induced ones.

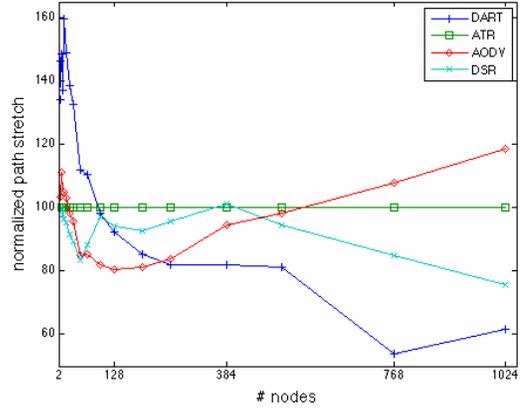

Figure 6 – Normalized path stretch

The last metric, the mean routing overhead normalized to the one of ATR protocol, measures the ability of a protocol of working well in congested or low-bandwidth environments. Let us note that, for DSR and AODV simulations, we count routing packets sent over multiple hops as a single transmission, as usually done, also if DART and ATR use only locally broadcasted routing packets. Fig.7 points out that the routing overhead of DART and ATR are perfectly comparable, that is the packet delivery performance gain of ATR is obtained with no additional overhead, with the exception of memory requirements. With respect to reactive protocols, ATR suffers from very high overhead for small networks. Whereas, when the number of nodes grows, the hierarchical routing overhead becomes perfectly comparable with the reactive one, wasting the aim of reactive protocols.

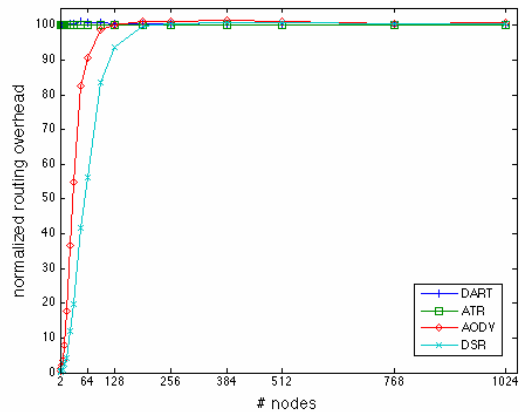

Figure 7 - Normalized routing overhead

### C. Address Allocation Process

To evaluate the improvements of our Address Allocation Process with respect to DART one, we set up a set of experiments with static topologies, no data traffic, simulation time equal to 450 seconds and nodes uniformly distributed. We measure the average times of last duplicate address and last invalid address event, for all the participating nodes. As shown in Fig. 8, the DART address allocation procedure never converges when the number of nodes is mode than 32.

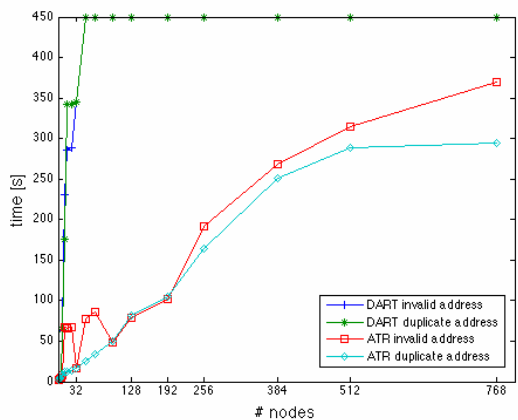

Figure 8 - Address Allocation Process convergence

*D. Address Lookup Process*

To analyze the overhead due to the address lookup and update functions, we have chosen two metrics: the mean rate of network address updates and the mean number of network address bits that have been changed. The former measures the temporal locality of the network addresses; the latter measures the spatial locality of the network address, and should be used to define the caching techniques.

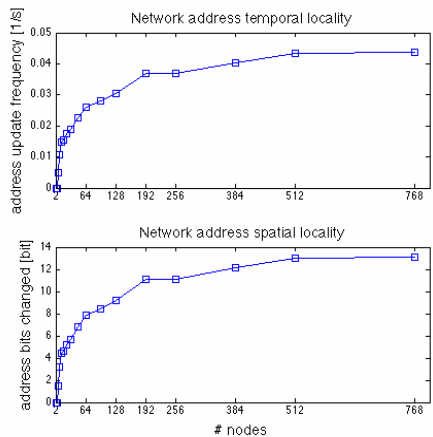

Figure 9 - Address Lookup Process metrics

The mean address-update rate is measured in updating per second and its trend initially grows more then linearly. However, when the number of nodes grows up, it is going to steady. The maximum value is in the order of an updating every 20 seconds, that is almost an order of magnitude higher than the time period necessary to route a data packet. Regards to the spatial locality metric, the high number of changed bits suggests to adopt distributed caching techniques.

## V. CONCLUSION

The paper proposes a hierarchical multi-path routing protocol, referred to as Augmented Tree-based Routing (ATR) protocol, which exploits a new augmented tree-based address-space structure, in order to solve the scalability problem and to gain good resilience against node failure/mobility and link congestion/instability in MANETs. Simulation results and performance comparisons with existing protocols substantiate the effectiveness of the ATR.

REFERENCES

[1] X. Hong, K. Xu and M. Gerla. "*Scalable routing protocols for mobile ad hoc networks Network*". IEEE Network, vol. 16, no. 4, 2002, pp. 11-21.
[2] J. Broch, D. A. Maltz, D. B. Johnson, Y. Hu and J. A. Jetcheva. "*A performance comparison of multi-hop wireless ad hoc network routing protocols*". In MobiCom '98: Proceedings of the 4th annual ACM/IEEE international conference on Mobile computing and networking, Dallas, Texas, United States,1998, pp. 85-97.
[3] Y. Tseng, S. Ni, Y. Chen and J. Sheu. "*The broadcast storm problem in a mobile ad hoc network*". Wireless Networks, vol. 8, no. 2, 2002, pp. 153-167.
[4] I. Chlamtac, M. Conti and J. Liu. "*Mobile ad hoc networking: imperatives and challenges*". Ad Hoc Networks, vol. 1, no. 1, 2003.
[5] I. Akyildiz, X. Wang and W. Wang. "*Wireless mesh networks: a survey*". Computer Networks, vol. 47, no. 4, 2005, pp. 445-487.
[6] M. Gerla, X. Hong and G. Pei. "*Landmark routing in ad hoc networks with mobile backbones*". J. of Parallel and Distributed Computing, vol. 63, no. 2, 2003, pp. 110-122.
[7] B. Chen and R. Morris. "*L+: Scalable landmark routing and address lookup for multi-hop wireless networks*". Tech. rep., Massachusetts Institute of Technology, MIT LCS-TR-837, 2002.
[8] J. Eriksson, M. Faloutsos and S. Krishnamurthy. "*Peernet: Pushing peer-2-peer down the stack*". In Proc. of IPTPS, 2003.
[9] C. Perkins and E. Royer. "*Ad hoc on-demand distance vector routing*". 2nd IEEE Workshop on Mobile Computing Systems and Applications, New Orleans, LA, United States, 1999, pp. 90-100.
[10] D. B. Johnson and D. A. Maltz. "*Dynamic source routing in ad hoc wireless networks*". In Mobile Computing, vol. 353, 1996, pp. 153-181.
[11] J. I. Alvarez-Hamelin, A. C. Viana and M. D. De Amorim. "*Architectural Considerations for a Self-Configuring Routing Scheme for Spontaneous Networks*". Page 1. arXiv:cs.NI/0510082, vol.1, 2005.
[12] J. Eriksson, M. Faloutsos and S. Krishnamurthy. "*DART: Dynamic Address RouTing for Scalable Ad Hoc and Mesh Networks*". IEEE-ACM Transactions on Networking, vol.15, no. 1, 2007, pp.119-132.
[13] S. Nesargi and R. Prakash. "*MANETconf: Configuration of hosts in a mobile ad hoc network*". In Proc. of IEEE INFOCOM, 2002.
[14] H. Zhou, L. M. Ni and M. W. Mutka, "*Prophet address allocation for large scale MANETs,*". Ad Hoc Networks, vol. 1, no. 4, 2003.
[15] M. Caesar, M. Castro, E. B. Nightingale, G. O'Shea and A. Rowstron. "*Virtual ring routing: network routing inspired by DHTs*". In Proc. of SIGCOMM, 2007.
[16] M. Caleffi, G. Ferraiuolo, and L. Paura. "On reliability of dynamic addressing routing protocols in mobile ad hoc networks. In proc of WRECOM '07, Rome, Italy, 2007.
[17] C. Bettstetter, "*On the minimum node degree and connectivity of a wireless multihop network*". In Proc. of ACM International Symposium on Mobile Ad Hoc Networking & Computing, pp.80–91, 2002.
[18] J. Yoon, M. Liu and B. Noble. "*Random waypoint considered harmful*". In Proceedings of IEEE INFOCOM 2003, vol.2, pp. 1312-1321, 2003.
[19] G. Holland and N. Vaidya. "*Analysis of TCP performance over mobile ad hoc networks*". In Proc. of the ACM/IEEE international conference on Mobile computing and networking, pp.219-230, 1999.
[20] J. Li, C. Blak, D. S. J. De Couto, H. I. Lee and R. Morris. "*Capacity of Ad Hoc wireless networks*". In Proc. of the international conference on Mobile computing and networking, pp.61-69, 2001.
[21] The VINT Project. The ns Manual (formerly ns Notes and documentation). Available at http://www.isi.edu/nsnam/ns/doc.
[22] D. Karger, E. Lehman, T. Leighton, M. Levine, D. Lewin, and R. Panigrahy, "*Consistent hashing and random trees: Distributed caching protocols for relieving hot spots on the world wide web*". On Proc. ACM Symp. Theory of Computing, pp.654-663, 1997.